\documentclass[
  aps,--
  pra,        
  amsmath,amssymb,
  reprint,  groupedaddress 
 ]{revtex4-2}

\usepackage[utf8]{inputenc}
\usepackage[T1]{fontenc}
\usepackage{mathptmx,soul}
\usepackage{comment}

\usepackage{graphicx}
\usepackage{dcolumn}
\usepackage{bm}
\usepackage{tikz}
\usetikzlibrary{decorations.pathmorphing, positioning, shapes.geometric, arrows.meta}

\usetikzlibrary{decorations.pathmorphing, positioning, shapes.geometric, arrows.meta}
\usepackage[section]{placeins}
\usepackage{url}
\usepackage{float}
\usepackage{etoolbox}

\usepackage{balance}

\makeatletter
\def\@email#1#2{%
 \endgroup
 \patchcmd{\titleblock@produce}
  {\frontmatter@RRAPformat}
  {\frontmatter@RRAPformat{\produce@RRAP{*#1\href{mailto:#2}{#2}}}\frontmatter@RRAPformat}
  {}{}
}%
\makeatother

\begin{document}

\title[Geometric Structure of Bell Correlations in Bohmian Mechanics: A Configuration-Space Analysis of EPR Experiments]%
      {Geometric Structure of Bell Correlations in Bohmian Mechanics: A Configuration-Space Analysis of EPR Experiments}

\author{Tim Dartois}
\author{Signe Seidelin}
\author{Aurélien Drezet} 
\affiliation{Univ. Grenoble Alpes, CNRS, Grenoble INP and Institut N\' eel, 38000 Grenoble, France}

\date{\today}

\begin{abstract}
 We develop an explicit configuration-space formulation of EPR-Bell experiments within the framework of de Broglie-Bohm theory, in which joint measurement outcomes arise from a deterministic mapping from initial particle configurations to outcome pairs. This construction induces a partition of the hidden-variable configuration space into domains associated with the different measurement outcomes. Using a reduced-dimensional Stern-Gerlach model, we derive the structure of these domains and identify the corresponding separatrices that define their boundaries. We show that Bell correlations emerge from the geometry of these partitions: the domain boundaries depend nonlocally on the measurement settings, while the marginal outcome distributions remain invariant, providing a direct dynamical realization of no-signaling. Analytical results are supported by numerical simulations, which exhibit quantitative agreement with the predicted domain structure as a consequence of the underlying partition of configuration space induced by the measurement dynamics. This approach provides an explicit configuration-space representation of nonlocal correlations in Bohmian mechanics, linking trajectory dynamics, measurement processes, and statistical predictions within a unified framework.

\end{abstract}

\maketitle

\section{Introduction}


The twentieth century marked the advent of quantum mechanics, now regarded as the most accurate physical theory for predicting microscopic phenomena. While its experimental successes are indisputable, its interpretation continues to fuel contemporary debates \cite{Laloe}. One of the central issues in these debates relies in a contradiction between two levels of description : on the one hand, there is the manifestly realistic and local nature of the macroscopic world; on the other hand, the microscopic world as described by quantum mechanics, where these notions of realism and locality seem to be called into question. 

Here, \emph{locality} refers to the principle that the outcome of a measurement performed on a physical system cannot be influenced by actions or settings applied to a distant system outside its past light cone. In other words, in a local theory, all physical influences are mediated by signals propagating at finite speed, and measurement results are determined solely by local variables and interactions. By \emph{realism} we mean the assumption that physical systems possess objective properties that exist independently of observation, so that measurement outcomes are determined by the physical state of the system prior to measurement, possibly in a context-dependent way. Since the 1920s, this paradox has been a source of tension in the scientific community, and has lead to a division between scientists. On one side are the realists, who support the existence of hidden variables that provide an underlying determinism to the quantum formalism; on the other side, the proponents of the orthodox interpretation, who accept a fundamentally indeterministic character.

These debates are at the heart of the 1927 Solvay conference where Louis de Broglie proposed the first "realistic" interpretation of quantum mechanics, describing the deterministic guidance of particles \cite{debroglie1928}. Known as pilot-wave theory, his proposal was rejected by the champions of the "antirealist" and  indeterministic orthodox interpretation (i.e., Bohr, Born, Heisenberg, Dirac, and Pauli), and in 1928 de Broglie himself eventually abandoned his realistic theory for a quarter of a century.

Despite this clear direction taken by the community, questions relating to the interpretation of quantum mechanics remained unresolved. In particular, Einstein remained convinced that quantum mechanics, despite its immense effectiveness, is not a “complete” theory in the sense that it should one day be superseded by a theory that restores realism. In 1935, Einstein, Podolsky, and Rosen (EPR) published an article that brought new elements to the discussion \cite{epr}. Using a pair of entangled quantum particles, EPR demonstrated a contradiction between the assumption that quantum mechanics is complete and the theory of special relativity, which assumes a principle of locality rejecting action-at-a-distance. Therefore, after EPR, we must admit that either quantum mechanics is local but incomplete, or that the theory is non-local. Einstein rejected the second alternative, which would lead to `spooky' action at a distance. Moreover, in 1952, the American physicist David Bohm independently rediscovered the pilot-wave theory first proposed 25 years earlier by de Broglie. He extended the theory and showed that it reproduces all statistical predictions of non-relativistic quantum mechanics \cite{bohm1952a} (a detailed review of the theory can be found in the literature \cite{Holland,DurrTeufel,DurrLazarovici,Norsenbook}). Importantly, de Broglie-Bohm  theory satisfies the EPR theorem \cite{deBroglie1951,bohm1952b} but is explicitly non-local, contradicting Einstein's desideratum.

Taking both the EPR article and Bohm's updated pilot-wave theory seriously, John Bell wondered whether the nonlocality of the de Broglie-Bohm theory, hereafter referred to as dBB \cite{note1}, was necessary.  In 1964 \cite{bell1964}, using a pair of entangled spin-1/2 particles forming a spin singlet as suggested by Bohm \cite{Bohm1951}, he was able to pose this problem in the form of an inequality that must necessarily be satisfied by any local theory - the precise term later used by Bell is  `locally-causal' theory \cite{Bellloccaus,Bellcitation}. Bell was followed by Clauser, Horne, Shimony, and Holt (CHSH) in 1969, who reformulated the theorem in a form suitable for experimental tests \cite{chsh1969}. Bell and CHSH showed that, under certain circumstances, quantum mechanics predicts correlations of the joint spin measurements of the two entangled particles violating their inequalities. In the words of Bell we can therefore conclude: `quantum mechanics cannot be embedded in a locally-causal theory' \cite{Bellcitation}.

The experimental violation of Bell inequalities was first demonstrated in the early 1970s by Freedman and Clauser \cite{freedman1972}, and later confirmed and significantly refined by a series of experiments, most notably those of Aspect and collaborators \cite{aspect} in the early 1980s. In modern experiments the entangled systems are typically pairs of photons, with polarization playing the role of the two-level spin degree of freedom.  Quantum mechanical predictions were thereby confirmed, while any locally-causal description of the underlying microscopic dynamics was ultimately ruled out \cite{note3}. In 2022, the Nobel Prize in Physics was finally awarded to  Clauser, Aspect, and Zeilinger for their pioneering experimental works proving quantum nonlocality. It is important to stress that this experimental confirmation does not rule out determinism and nonlocal hidden variables \cite{note2}. Thus, dBB theory, which is deterministic, features hidden variables and is explicitly nonlocal, remains a very serious candidate as a formulation of quantum physics.

In this context, it is quite remarkable to note that dBB is universally cited as an example of a nonlocal theory that violates Bell's inequalities. In particular, it provides a concrete illustration of how a theory \textbf{free of the measurement problem}, such as pilot-wave theory, can nevertheless account for the Bell correlations predicted by standard quantum mechanics. Much work \cite{remark} has focused on addressing the EPR problem using entangled spin-1/2 particles \cite{dewdney1987,dewdney1988,norsen2014,gondran2016,dewdney2023}. In parallel, other works have sought to probe Bohmian nonlocality experimentally without relying directly on Bell-type inequalities, in particular through weak-measurement protocols \cite{Kocsis2011,Mahler2016,Dou2025,Wang2025}. These studies reconstruct average trajectories or trajectory-based nonlocal effects in interferometric settings, thereby illustrating the usefulness of Bohmian dynamical concepts for analyzing quantum phenomena (see also \cite{Elsayed2018,Foo2022,Sunkyu2025,Grande2026,Sharoglazova2025,Drezet2025}). Moreover, such results should be understood as supporting the formal and operational relevance of Bohmian-like structures, rather than establishing the ontological validity of de Broglie-Bohm theory.  
However, an explicit configuration-space formulation of Bell-type experiments within the dBB framework, in which joint measurement outcomes are directly linked to the underlying hidden-variable configuration dynamics, remains largely undeveloped. Moreover, a statistical analysis formulated directly in hidden-variable configuration space, in the spirit of Bell’s original construction, is still missing in an explicit configuration-space representation. This provides a direct structural account of Bell correlations within the deterministic dynamics of the theory.
From a practical perspective, a deeper understanding of dBB trajectories can provide valuable guidance for the design and interpretation of new experiments \cite{Kocsis2011,Mahler2016,Dou2025,Wang2025}.

In this work, we show that Bell correlations in Bohmian mechanics arise from the structure of partitions of hidden-variable configuration space induced by the measurement dynamics. We develop an explicit configuration-space construction of an EPR-Bell-type experiment within the dBB framework, implemented analytically and numerically using magnetic coils to model local spin rotations. These coils, acting as spin-flippers \cite{spinflip1}, allow us to avoid rotating the massive Stern-Gerlach analyzers themselves. In this way, we remain consistent with the essentially one-dimensional character of the experiment as analyzed in earlier contributions \cite{dewdney1987,dewdney1988,norsen2014,gondran2016,dewdney2023}.

The present article is organized as follows. First, we analyze the Stern-Gerlach experiment, which serves as a reminder of pilot-wave theory and the deterministic interpretation it offers of quantum mechanics. We then construct the corresponding dynamical mapping from initial conditions to measurement outcomes for a Bell-type experiment. We sample the initial positions of entangled particles in a Bell state according to the modulus-squared of the wave function and evolve the associated wavefunctions in time, analyzing the resulting trajectory dynamics. Next, we analyze the induced partition of hidden-variable configuration space and the associated outcome statistics, leading to the emergence of Bell-inequality-violating correlations. This construction shows explicitly how a deterministic and nonlocal theory gives rise to Bell-inequality-violating correlations while describing definite particle trajectories. Finally, we show how setting-independent marginal statistics emerge from a dynamics whose outcome structure depends nonlocally on distant measurement settings. To this end, we construct explicit representations of the hidden-variable configuration space, revealing the partition into outcome domains and the geometry of their boundaries. Within this framework, entanglement correlations emerge directly from the dynamical partition of hidden-variable space induced by the measurement settings.

\section{Description of the experiment}

\subsection{Overview and approach}

The construction we implement is inspired by Bell’s original thought experiment \cite{bell1964} and provides a dynamical realization of an EPR–Bell scenario within the Broglie-Bohm framework, schematically summarized in Fig.~\ref{schemapcp}. The setup consists of a source emitting two spin-entangled particles in opposite directions toward two spatially separated observers, Alice and Bob, each equipped with a Stern-Gerlach apparatus. The two apparatuses are separated by an arbitrarily large distance. Alice’s Stern-Gerlach apparatus implements a spin measurement by coupling the particle to a magnetic field gradient, leading to spatial separation of spin components from which she can infer the spin outcome. Bob performs an analogous measurement on the second particle after a controlled time delay relative to Alice’s measurement. This temporal ordering simplifies the analysis of the dynamical evolution and allows a clear identification of the dependence of Bob’s outcome on the measurement configuration.

As mentioned, we choose to model this experiment in one dimension. In Fig.~\ref{schemapcp} the horizontal direction represents time: the particles are emitted at the center (early times), and their subsequent evolution is shown as they propagate outward on both sides toward later times. In a genuine three-dimensional experiment, the particles would be physically sent toward the Stern-Gerlach devices. In our model, the particles are initially localized and then subjected to the magnetic field gradient. The reduced-dimensional model preserves the essential structure of the measurement dynamics while allowing a transparent analysis of the underlying Bohmian configuration-space dynamics.

The main difference between Bell's original experiment and our model -as well as in other works \cite{dewdney1987,dewdney1988,norsen2014,gondran2016,dewdney2023}- lies in how the experimental parameters can vary. In Bell's experiment, the measurement parameters are the orientations of the Stern-Gerlach devices.  To remain consistent with the one-dimensional formulation of the model, the measurement settings are encoded through local spin rotations implemented by magnetic coils acting on each branch of the experiment. We introduce Bell-type parameters that control the measurement configuration and reveal the dependence of the hidden-variable dynamics on distant settings. This construction closely parallels Bell’s original formulation, while making the dependence of the outcome structure on hidden-variable space explicit. To construct the dBB description of this experiment, we first formulate the Stern-Gerlach measurement as a deterministic mapping in configuration-space for individual particles. The full Bell experiment is then obtained by combining two such mappings, yielding a joint dynamical evolution in hidden-variable space.
 
\begin{figure*}
    \centering
        \begin{tikzpicture}[>=Stealth, every node/.style={font=\small}]
      \useasboundingbox (-7,-3.5) rectangle (7,3.2);

      \node[circle, draw, minimum size=14mm, inner sep=1mm] (S) at (0,0) {$S$};

      \draw[thick] (S.west) -- (-4,0)
        node[midway, above=4mm] {$e^-$ Alice};
      \draw[thick] (S.east) -- (4,0)
        node[midway, above=4mm] {$e^-$ Bob};

        \begin{scope}[shift={(-4,0)}]
            \begin{scope}[shift={(0,0.35)}]
                \draw[thick]
                  (-0.3,0) -- (0.3,0) -- (0.6,1.5) -- (-0.6,1.5) -- cycle;
            \end{scope}
        
            \begin{scope}[shift={(0,-0.35)}]
                \draw[thick]
                  (-0.3,0) -- (0.3,0) -- (0.6,-1.5) -- (-0.6,-1.5) -- cycle;
            \end{scope}
        
            \draw[->,thick] (0,0) -- (-1.6,1.2) node[left] {$+$ (up)};
            \draw[->,thick] (0,0) -- (-1.6,-1.2) node[left] {$-$ (down)};
            \node at (0,-2.35) {SG$_z$--Alice};
            \draw[->,thick, dashed] (0.0,-1.2)--(0,1.2)
              node[above] {$-\vec{\nabla} B_z$};
        \end{scope}

        \begin{scope}[shift={(4,0)}]
            \begin{scope}[shift={(0,0.35)}]
                \draw[thick]
                  (-0.3,0) -- (0.3,0) -- (0.6,1.5) -- (-0.6,1.5) -- cycle;
            \end{scope}
        
            \begin{scope}[shift={(0,-0.35)}]
                \draw[thick]
                  (-0.3,0) -- (0.3,0) -- (0.6,-1.5) -- (-0.6,-1.5) -- cycle;
            \end{scope}
        
            \draw[->,thick] (0,0) -- (1.6,1.2) node[right] {$+$ (up)};
            \draw[->,thick] (0,0) -- (1.6,-1.2) node[right] {$-$ (down)};
            \node at (0,-2.35) {SG$_z$--Bob};
            \draw[->,thick, dashed] (0.0,-1.2)--(0,1.2)
              node[above] {$-\vec{\nabla} B_z$};
        \end{scope}

      \draw[decorate, decoration={coil, segment length=6pt, amplitude=3pt}, thick]
        (1.6,0) -- (2.8,0);
      \node at (2.2,-0.5) {Coil--Bob $(\beta)$};

      \draw[decorate, decoration={coil, segment length=6pt, amplitude=3pt}, thick]
        (-2.8,0) -- (-1.6,0);
      \node at (-2.2,-0.5) {Coil--Alice $(\alpha)$};

      \begin{scope}[shift={(0,-2.4)}]
        \draw[thick] (8.2,0) -- (7,0) node[left=0mm] {$y$};
        \draw[thick] (8.2,0) -- (8.2,1.2) node[above] {$z$};
      \end{scope}--
    \end{tikzpicture}
    \caption{\label{schemapcp}
    Schematic of the EPR-Bell experiment considered in this work. A source $S$ emits two spin-entangled particles prepared in a singlet state and propagating toward two distant observers, Alice (left) and Bob (right). Each particle passes through a local magnetic coil, which implements a spin rotation by an angle $\alpha$ (Alice) or $\beta$ (Bob) about the horizontal axis, thereby encoding the measurement settings. The Stern-Gerlach analyzers are fixed and oriented along the $z$ direction, so that spin information is mapped onto a spatial separation along $z$. The protocol is explicitly time-ordered: Alice’s particle traverses its coil and Stern-Gerlach apparatus and is measured before Bob’s particle reaches his analyzer. This ordering allows a clear identification of the dependence of Bob’s outcome on the distant measurement configuration while preserving the full nonlocal structure of the Bohmian dynamics. The coils thus play the role of effective measurement settings in Bell’s scenario, replacing physical rotations of the Stern-Gerlach apparatus and preserving the essentially one-dimensional formulation of the model.
    }
\end{figure*}

\subsection{Spin and Stern-Gerlach experiment in pilot-wave theory}\label{sgmodel}

We now briefly recall the formulation of de Broglie-Bohm (dBB) theory relevant for the Stern-Gerlach experiment, which will serve as the basic building block of our numerical model \cite{Holland,DurrTeufel,DurrLazarovici}. 

In the non-relativistic approximation, the dynamics of a spin-$\frac{1}{2}$ particle in an external electromagnetic field is described by the Schrödinger-Pauli equation \cite{Bellbook}
\begin{equation}
i\hbar\frac{\partial \boldsymbol{\Psi}}{\partial t}=H\boldsymbol{\Psi},
\end{equation}
with
\begin{equation*}
H = \frac{-\hbar^2}{2m}\left[ \vec{\nabla}-\frac{ie}{\hbar c}\vec{A} \right]^2-\hat{\vec{\mu}}\cdot\vec{B} +eV,
\qquad
\hat{\vec{\mu}}=\gamma\frac{\hbar}{2}\hat{\vec{\sigma}},
\end{equation*}
where $\hat{\vec{\sigma}} := (\hat{\sigma}_{x},\hat{\sigma}_{y},\hat{\sigma}_{z})$ is the vector of Pauli matrices, and $\gamma:=\frac{ge}{2m}$ is the gyromagnetic ratio, with $g$ the Landé factor and $e$ the particle charge. 
The wave function $\boldsymbol{\Psi}$ is a two-component spinor, $\boldsymbol{\Psi}=(\Psi_{+},\Psi_{-})$, corresponding to the positive and negative spin components $\pm \frac{\hbar}{2}\vec{e}_z$ along the $z$ direction.

Repeating the usual reasoning applied to the Schrödinger equation, one obtains a continuity equation for the probability density
\begin{equation}
\rho(\vec r,t)=\boldsymbol{\Psi}^\dagger\boldsymbol{\Psi},
\end{equation}
and a conserved probability current of the form
\begin{equation}\label{courantspin}
\vec{j}=\frac{\hbar}{m} \Im(\boldsymbol{\Psi}^\dagger\vec{\nabla}\boldsymbol{\Psi})
-\frac{e}{mc}\vec{A}\boldsymbol{\Psi}^\dagger\boldsymbol{\Psi}
+\frac{\hbar}{2m}\vec{\nabla}\times(\boldsymbol{\Psi}^\dagger\hat{\vec{\sigma}}\boldsymbol{\Psi}),
\end{equation}
which satisfies
\begin{equation}
\partial_t \rho + \vec{\nabla}\cdot\vec{j}=0.
\end{equation}

The main point of dBB theory is to interpret the current $\vec{j}(\vec r,t)$ as a material current field. One then introduces the particle position $\vec r(t)$ and defines the particle velocity through the ratio of the current to the density evaluated at the particle position:
\begin{equation}\label{guidanceeq}
\vec{v}(\vec{r},t)=\frac{d\vec{r}(t)}{dt}
=\frac{\vec{j}(\vec{r},t)}{\rho(\vec{r},t)}\Big|_{\vec{r}(t)} .
\end{equation}
This relation defines a deterministic dynamics in which particle trajectories are guided by the spinor wave function $\boldsymbol{\Psi}$.

We see that in the expression \eqref{courantspin} for the current there appears a rotational term (the curl term). This term is generally omitted in Bohmian studies of spin dynamics. The omission is ``allowed'' because the continuity equation only involves the divergence of the current; thus equivariance is guaranteed even if one uses a current without the rotational part. In other words, whether one includes or omits the rotational term in the velocity, the statistical predictions of pilot-wave theory remain identical to those of quantum mechanics \cite{warning}. In this article, following Bell \cite{Bellrmp} himself and most works \cite{dewdney1987,dewdney1988,norsen2014,gondran2016,dewdney2023}, we also omit the rotational term in order to remain consistent with our one-dimensional model.

It is important to state an essential assumption. In the approach presented here we implicitly assume the hypothesis of \emph{quantum equilibrium} \cite{Valentinia,Valentinib}, according to which the statistical distribution of initial positions satisfies
\begin{equation}
\rho(\vec r,0)=|\boldsymbol{\Psi}(\vec r,0)|^2.
\end{equation}
This condition, preserved by the continuity equation, guarantees that the theory reproduces exactly the statistical predictions of non-relativistic quantum mechanics \cite{bohm1952a,bohm1952b,Holland,DurrTeufel,DurrLazarovici}.

We now turn to the Stern-Gerlach experiment. In this setup, a beam of spin-$\frac{1}{2}$ particles passes through an inhomogeneous magnetic field, so that the spin-field interaction separates the outgoing wave packet into two spatially distinct components associated with the eigenvalues of the spin projection along the field direction. The particles are then deflected toward opposite regions of the detection screen, allowing the spin outcome to be inferred from the sign of the deflection. In the present framework, the Stern-Gerlach device plays a central role as a concrete dynamical realization of spin measurement: it implements a mapping from the spinor wave function to spatially separated wave packets, which can be directly analyzed within pilot-wave theory through the guidance equation (Eq. \ref{guidanceeq}).

Since the pioneering work of Bohm, Schiller and Tiomno in 1955 \cite{bohm1955A,bohm1955B}, up to more modern formulations such as those given by Dewdney \cite{dewdney1988}, detailed studies of spin correlations in EPR-type experiments \cite{dewdney1987} and contemporary summaries \cite{norsen2014}, a clear and deterministic description of spin in the Bohmian framework has emerged. These results provide the theoretical basis for modeling the Stern-Gerlach dynamics and, ultimately, for constructing the EPR-Bell experiment considered here.

To construct the Stern-Gerlach measurement within pilot-wave theory, we follow the approach of Dewdney, Holland, and Kyprianidis \cite{dewdney1986PLA}. We consider a one-dimensional dynamics, in which the wave packet propagates along the $z$ axis while evolving in time. The incoming state is taken to be a Gaussian wave packet, describing free evolution upstream of the magnetic field region. The interaction with the magnetic field is modeled as a strong and short impulse of duration $T$, whose detailed treatment is given in Sec. II.C.d. In this regime, the Stern-Gerlach device implements a dynamical mapping from initial wavefunction and particle position to spatially separated components associated with the measurement outcomes. In this approximation, the spatial separation of the wave packet develops only after the interaction region.

We model the magnetic field as $ B=B_{0}+zB_{0}'$. 
In this regime, the outgoing wave function has two spin components:
\begin{equation}\label{spinsg1}
\boldsymbol{\Psi}(z,t)=c_+\boldsymbol{\Psi}_{+}(z,t)+c_-\boldsymbol{\Psi}_{-}(z,t),
\end{equation}
and these components admit the analytical form :
\begin{equation}
\Psi_{\pm}(z,t)=(2\pi s_{t})\cdot \exp\!\left[- \frac{(z\mp ut)^2}{4 \sigma_{0} s_{t}}\pm i\big(\Delta+(z\mp\tfrac{1}{2}ut)\Delta '\big)\right],
\end{equation}
where $\Delta:=\frac{\mu B_{0} T}{\hbar}$, $\Delta':=\frac{\mu B_{0}' T}{\hbar}$, $u=\frac{\hbar\Delta'}{m}$ is the ``velocity'' of the center of the wave packet, and $s_{t}=\sigma_{0}\bigl(1+\frac{i\hbar t}{2m\sigma_{0}^2}\bigr)$ is the time-dependent width of the packet. The coefficients $c_{\pm}$ determine how the spinor wave function is distributed among the $\pm$ spin states, and satisfy $|c_{+}|^2+|c_{-}|^2 = 1$.

The structure of the wave function shows that it decomposes into two Gaussian wave packets associated with the two spin components. For each component ($\pm$), the center of the corresponding wave packet propagates with velocity $\pm u$. Within the Bohmian framework, particle trajectories are determined by the guidance equation, so that this decomposition induces a separation of trajectories into two distinct bundles associated with the spinor components. This mechanism underlies the dynamical mapping between initial conditions and measurement outcomes, which is the point of view adopted by Bell and by Dewdney and Holland.

\subsection{From the Stern-Gerlach experiment to an EPR-Bell setup}

We now construct the EPR-Bell configuration by combining two Stern-Gerlach measurement processes through an entangled initial state. This defines a joint dynamical mapping from the initial hidden variables of the two particles to the pair of measurement outcomes. Local spin rotations and sequential measurements provide a concrete realization of the measurement settings entering Bell’s original formulation. The extension from the single-particle Stern-Gerlach dynamics to the two-particle case follows naturally by considering an entangled initial state. We focus on the antisymmetric Bell state relevant to our construction:

\begin{equation}\label{etatbell}
\boldsymbol{\Psi}(\vec{r}_{1},\vec{r}_{2},t)=\frac{1}{\sqrt{2}}\left(\boldsymbol{\Psi}_{+}(\vec{r}_{1},t)\boldsymbol{\Psi}_{-}(\vec{r}_{2},t)-\boldsymbol{\Psi}_{-}(\vec{r}_{1},t)\boldsymbol{\Psi}_{+}(\vec{r}_{2},t)\right),
\end{equation}
where $\vec{r}_{n}, n=1,2$ is the position variable of particle $n$ and $\boldsymbol{\Psi}_{\pm}(\vec{r}_{n},t)$ is a spinor spin eigenstate for the $\pm$ spin component along $z$ of the $n$th particle.\\
\indent In the general case of a multi-particle dynamics with identical masses, the density and the current associated with the motion of the $n$th particle are defined by
\begin{equation}\label{rhobell}
\rho=\boldsymbol{\Psi}^\dagger\boldsymbol{\Psi},\qquad\vec{j}_{n}=\frac{\hbar}{m}\Im(\boldsymbol{\Psi}^\dagger\vec{\nabla}_{\vec{r}_{n}}\boldsymbol{\Psi}).
\end{equation}
Note that such an expression is only valid in the absence of a vector potential ($\vec{A}=0$).
The velocity of the $n$th particle is then given by
\begin{equation}
\frac{d}{dt}\vec{R}_n(t)=\vec{v}_{n}=\left.\frac{\vec{j}_{n}}{\rho}\right|_{(\vec{r_{1}},\vec{r_{2}})=(\vec{R_{1}}(t),\vec{R_{2}}(t))},
\end{equation}
with $(\vec{R_{1}}(t),\vec{R_{2}}(t))$ the positions of particles $(1,2)$ at time $t$.\\
\indent In the particular case of the Bell state \eqref{etatbell}, the currents and the density \eqref{rhobell} become
\begin{equation}
\begin{aligned}
\vec{j}_{1/2}(\vec r_1,\vec r_2,t)
= \frac{\hbar}{m}\,\Im\Big[
&\,|\Psi_{2/1,+}|^2\,
\Psi^\dagger_{1/2,-}\,\vec\nabla_{1/2}\Psi_{1/2,-} \\
&+ |\Psi_{2/1,-}|^2\,
\Psi^\dagger_{1/2,+}\,\vec\nabla_{1/2}\Psi_{1/2,+}
\Big]
\end{aligned}
\end{equation}

\begin{equation}
\rho(\vec{r}_{1},\vec{r}_{2},t)=|\Psi_{1,+}|^2|\Psi_{2,-}|^2+|\Psi_{2,+}|^2|\Psi_{1,-}|^2,
\end{equation}
where $\Psi_{i,\pm}:=\Psi_{\pm}(\vec{r}_{i},t)$ and $\vec{\nabla}_{i}:=\vec{\nabla}_{\vec{r}_{i}}$.

Extending the numerical integration of the dynamics to two particles then amounts to solving two coupled first-order differential equations.

We now propose to decompose the dynamics into several stages, in order to remain as close as possible to Bell's experiment. First, the two particles are guided by one-dimensional Gaussian wave packets, which are free solutions of the Schrödinger equation and entangled in a Bell state \eqref{etatbell}. We choose the quantization axis of the Bell state \eqref{etatbell} to be the $z$ axis:
\begin{equation}\label{etatbellz}
\boldsymbol{\Psi}(z_{A},z_{B},t)=\frac{\boldsymbol{\Psi}_{+_{z}}(z_{A},t)\boldsymbol{\Psi}_{-_{z}}(z_{B},t)-\boldsymbol{\Psi}_{-_{z}}(z_{A},t)\boldsymbol{\Psi}_{+_{z}}(z_{B},t)}{\sqrt{2}},
\end{equation}
where $z_A$ (resp. $z_B$) denotes the coordinate of the particle sent to Alice (resp. Bob).

The two particles then pass, one after the other, through coils generating static and uniform magnetic fields oriented along directions $\vec{n}_{A}$ and $\vec{n}_{B}$, which form angles $\alpha$ and $\beta$ with the $z$ axis. The magnetic fields generated by the coils  implementing a spin rotation encoding the measurement settings. Particle $A$ then passes through a Stern-Gerlach apparatus oriented along $z$, while particle $B$ continues its free evolution. Later, particle $B$ passes through a Stern-Gerlach apparatus also oriented along $z$. The resulting dynamical mapping from initial conditions to outcomes defines, in hidden-variable configuration space, a partition into outcome domains whose statistical structure gives rise to the Bell correlations.
 
\paragraph{Free evolution}
The spatial wave packets are initially Gaussian and independent of the particle spin. As a result, the two-particle wave function \eqref{etatbellz} factorizes as

\begin{equation}
\boldsymbol{\Psi}_{1}(z_{A},z_{B},t)=\frac{1}{\sqrt{2}}G_{2}(z_{A},z_{B},t)\bigl(\left| +_{z},-_{z} \right\rangle-\left| -_{z},+_{z} \right\rangle\bigr),
\end{equation}
where $G_{2}(z_{A},z_{B},t) :=G(z_{A},t)G(z_{B},t)$ is the product of two Gaussian wave packets associated with each particle.\\

\paragraph{Passage of a particle through a `spin-flipper' \rm \cite{noteRauch}}

We represent the interaction between the magnetic field generated by the coils and the spin part of the wave function through the Hamiltonian:

\begin{equation}
\hat{H}_{mag}=-\hat{\vec{\mu}}\cdot\vec{B},\qquad\hat{\vec{\mu}}=\gamma\frac{\hbar}{2}\hat{\vec{\sigma}},
\end{equation}
We choose a magnetic field oriented along the $\vec{e}_y$ axis, $\vec{B}=B\,\vec{e}_y$. The interaction Hamiltonian then becomes
\begin{equation}
\hat{H}_{mag}=-\mu B\hat{\sigma}_{y}.
\end{equation}
Introducing $\boldsymbol{\zeta}$, a one-particle spinor wave function, the time evolution of $\boldsymbol{\zeta}(t)$ is given by
\begin{equation}
i\hbar\partial_t\boldsymbol{\zeta}(t)=\hat{H}_{mag}\boldsymbol{\zeta}(t)=-\mu B\hat{\sigma}_{y}\boldsymbol{\zeta}(t),
\end{equation}
and consequently
\begin{equation}
\boldsymbol{\zeta}(t+\Delta t)=e^{-i\frac{\hat{H}_{mag}}{\hbar}\Delta t}\boldsymbol{\zeta}(t)=e^{i\frac{\alpha}{2}\hat{\sigma}_{y}}\boldsymbol{\zeta}(t),\quad\alpha:=\frac{\mu B \Delta t}{\hbar}.
\end{equation}
Writing
\begin{equation}
\boldsymbol{\zeta}(t)= \begin{pmatrix}
 u\\
 v
\end{pmatrix}_{z},
\end{equation}
with arbitrary $u$ and $v$, and using
\begin{equation}
e^{i\frac{\alpha}{2}\hat{\sigma}_{y}}=\cos(\tfrac{\alpha}{2})\hat{\boldsymbol{1}}+i\sin(\tfrac{\alpha}{2})\hat{\sigma}_{y},
\end{equation}
we obtain
\begin{equation}
\begin{aligned}
\boldsymbol{\zeta}(t+\Delta t)
&=(\cos(\tfrac{\alpha}{2})\hat{\boldsymbol{1}}
+i\sin(\tfrac{\alpha}{2})\hat{\sigma}_{y})\,\boldsymbol{\zeta}(t)  \\
&=\begin{pmatrix}
 \cos(\tfrac{\alpha}{2})u+\sin(\tfrac{\alpha}{2})v\\
 \cos(\tfrac{\alpha}{2})v-\sin(\tfrac{\alpha}{2})u
\end{pmatrix}_{z}
=\begin{pmatrix}
 u\\ v
\end{pmatrix}_{z'} .
\end{aligned}
\end{equation}
Here $z'$ is obtained by rotating the physical $z$ axis by an angle $\alpha$ around the $y$ axis.
The effect of this transformation on the basis vectors $\left| \pm_{z}\right\rangle$ is thus
\begin{equation}
\left\{
\begin{aligned}
\left| +_{z} \right\rangle &\mapsto \cos(\alpha/2)\left| +_{z} \right\rangle-\sin(\alpha/2)\left| -_{z} \right\rangle=\left| +_{z'} \right\rangle,\\
\left| -_{z} \right\rangle &\mapsto \cos(\alpha/2)\left| -_{z} \right\rangle+\sin(\alpha/2)\left| +_{z} \right\rangle=\left| -_{z'} \right\rangle.
\end{aligned}
\right.
\end{equation}
This transformation corresponds to a rotation of the spinor by an angle $\tfrac{\alpha}{2}$ about the $y$ axis. Such a rotation in spin space corresponds to a rotation of angle $\alpha$ in real space, i.e., in the $xOz$ plane. Applying rotations of angles $\alpha$ and $\beta$ to particles $A$ and $B$ respectively, the two-particle wave function becomes
\begin{equation}
\begin{aligned}
\boldsymbol{\Psi}_{2}(z_{A}, z_{B}, t)
&=\frac{G(z_{A},z_{B}, t)}{\sqrt{2}}
\Big[
\left| +_{z}\right\rangle
\bigl(\sin\tfrac{\gamma}{2}\left| +_{z}\right\rangle
      +\cos\tfrac{\gamma}{2}\left| -_{z}\right\rangle\bigr)
\\[4pt]
&\quad\;\;
+\left| -_{z}\right\rangle
\bigl(\sin\tfrac{\gamma}{2}\left| -_{z}\right\rangle
      -\cos\tfrac{\gamma}{2}\left| +_{z}\right\rangle\bigr)
\Big],
\end{aligned}
\end{equation}
where $\gamma:=\beta-\alpha$.\\

\paragraph{Passage of the particles through the Stern-Gerlach analyzers} 

After particle $A$ passes through its Stern-Gerlach apparatus, the spatial part of its wave function splits into two components, as described in Sec.~\ref{sgmodel}, inducing a corresponding separation in the trajectory dynamics. The other part of the wave function remains Gaussian. The Bell wave function after particle $A$ has passed through the Stern-Gerlach device is thus
\begin{equation}
\begin{aligned}
\boldsymbol{\Psi}_{3}(z_{A}, z_{B}, t)
&= \frac{G(z_{B}, t)}{\sqrt{2}}
\Big[
\boldsymbol{\Psi}_{A,+}\left| +_{z}\right\rangle
\bigl(\sin\tfrac{\gamma}{2}\left| +_{z}\right\rangle
    +\cos\tfrac{\gamma}{2}\left| -_{z}\right\rangle\bigr)
\\[4pt]
&\quad\;\;
+\boldsymbol{\Psi}_{A,-}\left| -_{z}\right\rangle
\bigl(\sin\tfrac{\gamma}{2}\left| -_{z}\right\rangle
    -\cos\tfrac{\gamma}{2}\left| +_{z}\right\rangle\bigr)
\Big],
\end{aligned}
\end{equation}
where $\boldsymbol{\Psi}_{A,\pm}:=\Psi_{\pm}(z,t)\left| \pm_{z}\right\rangle$ and $\Psi_{\pm}(z_A,t)$ is given by Eq.~(3). In the same way, when particle $B$ interacts with its Stern-Gerlach apparatus, the spatial part of its wave function splits analogously, yielding the final two-particle state 
\begin{equation}
\begin{aligned}
\boldsymbol{\Psi}_{4}
&=\boldsymbol{\Psi}_{A,+}\bigl(
 \cos(\tfrac{\gamma}{2})\,\boldsymbol{\Psi}_{B,-}
+\sin(\tfrac{\gamma}{2})\,\boldsymbol{\Psi}_{B,+}\bigr) \\
&\quad
+\boldsymbol{\Psi}_{A,-}\bigl(
 \sin(\tfrac{\gamma}{2})\,\boldsymbol{\Psi}_{B,-}
-\cos(\tfrac{\gamma}{2})\,\boldsymbol{\Psi}_{B,+}\bigr).
\end{aligned}
\end{equation}

The experiment described here is, up to minor differences, the Bohmian version of Bell's original experiment. The only slight change is that, in the original setup, the Stern-Gerlach apparatuses themselves are rotated, whereas in our case the particle spins are rotated by magnetic coils. The use of magnetic coils instead of rotating Stern-Gerlach analyzers allows the measurement settings to be implemented while preserving a one-dimensional dynamical description. This choice also provides a simpler and more controlled realization of the measurement configuration.

\paragraph{Intermediate regimes}\label{inter}

The wave functions associated with the different free-evolution regimes have been specified above. However, the so-called \emph{intermediate} regimes have not yet been discussed. In our setup, these regimes correspond to the time intervals during which the particles are subjected to the magnetic field gradients generated by the Stern-Gerlach devices over a short duration $T$. For particle $A$ (respectively particle $B$), this interaction regime takes place between times $t_1$ and $t_2$ (respectively $t_3$ and $t_4$). These intervals are defined such that $t_2 - t_1 = t_4 - t_3 = T$, and satisfy $t_1 \ll t_3$, meaning that Alice performs her measurement well before Bob. Finding exact analytical expressions for the wave functions during these intermediate regimes is extremely difficult. Rather than attempting to derive exact solutions, we approximate the transitions between these regimes. Although we do not model explicitly the detailed interaction of the particles with the magnetic field gradients, we impose continuity of both the probability density and the probability current in order to preserve quantum equilibrium. We therefore propose, for times between the beginning of the interaction at $t_N$ and its end at $t_{N+1}$, with $N = 1,3$, a smooth interpolation between the corresponding free regimes. For the probability current associated with particle $i = A,B$ during this transition, we define
\begin{equation}
\vec{j}_{\mathrm{int},N,i}(t)
=
\left(1 - \lambda_N(t)\right)\vec{j}_{N,i}(t)
+
\lambda_N(t)\vec{j}_{N+1,i}(t),
\end{equation}
where $\vec{j}_{K,i}(t)$ denotes the probability current associated with particle $i$ in the state $\boldsymbol{\Psi}_K(t)$, with $K = 1,\dots,4$.

The function $\lambda_N(t)$ is chosen as a smooth transition function given by
\begin{equation}
\lambda_N(t)
=
\frac{1 - \cos\!\left( \pi \frac{t - t_{N+1}}{t_N - t_{N+1}} \right)}{2}.
\end{equation}

The corresponding transition density is then naturally defined as
\begin{equation}
\rho_{\mathrm{int},N}(t)
=
\left(1 - \lambda_N(t)\right)\rho_N(t)
+
\lambda_N(t)\rho_{N+1}(t),
\end{equation}
where $\rho_K(t)$ denotes the probability density associated with the state $\boldsymbol{\Psi}_K(t)$.

Within these intermediate time intervals, the particle velocity remains defined as the ratio of the probability current to the probability density, evaluated at the particle position.


\section{Numerical analysis of the EPR-Bell pilot-wave model}
\subsection{Numerical method and sampling of initial conditions}

We analyze the Bohmian particle dynamics associated with the EPR-Bell configuration defined above. The wave functions fully determine the trajectory dynamics through the guidance equation, yielding a well-defined mapping from initial conditions to particle trajectories. Numerically, this mapping is obtained by integrating the equations of motion for an ensemble of initial conditions and examining the resulting trajectories.

In order for the Bohmian dynamics to agree with the statistical predictions of quantum mechanics, the initial conditions must be distributed according to the modulus-squared of the wave function. This condition enforces quantum equilibrium and ensures consistency with quantum mechanical statistics. The equations of motion are coupled first-order differential equations. We use a finite-difference scheme to integrate the dynamics. For sampling the initial conditions, we use the inversion method for the cumulative distribution function associated with the modulus-squared of the initial wave function. 

We employ the inversion method for sampling from the modulus-squared distribution of the initial wave function. This method consists in applying the inverse cumulative distribution function associated with a probability density $f$ to a uniformly distributed random variable on $(0,1)$.

For a distribution $f$, we define the associated cumulative function
\begin{equation}
F(x) := \int_{x_{min}}^{x} f(x') dx'.
\end{equation}
One can then show that if $X \sim \mathcal{U}(0,1)$, then $F^{-1}[X] \sim f$. In other words, if $X$ is uniformly distributed on $(0,1)$, then $F^{-1}[X]$ is distributed according to $f$.\\
The initial wave function has symmetries that reduce the sampling problem to a one-dimensional one. Indeed,
\begin{equation}
|\boldsymbol{\Psi}_{1}(z_{A},z_{B},0)|^2=\frac{1}{2\pi s^2_{0}}\exp\!\left(-\frac{z^2_{A}+z^2_{B}}{2s^2_{0}}\right),
\end{equation}
which can be written in polar coordinates as
\begin{equation}
|\boldsymbol{\Psi}_{1}'(\rho,\theta,0)|^2=\frac{1}{2\pi s^2_{0}}\exp\!\left(-\frac{\rho^2}{2s^2_{0}}\right),
\end{equation}
with $\rho := \sqrt{z^2_{A}+z^2_{B}}$. This function does not depend on $\theta$ and can be ``uniformized'' as
\begin{equation}
|\boldsymbol{\Psi}_{1}''|^2(\rho) :=\int_{0}^{2\pi}|\boldsymbol{\Psi}_{I}'(\rho,\theta,0)|^2d\theta=\frac{1}{s^2_{0}}\exp\!\left(-\frac{\rho^2}{2s^2_{0}}\right).
\end{equation}
We can thus extract random variables $(Z_{A},Z_{B})$ distributed according to $|\boldsymbol{\Psi}_{1}(z_{A},z_{B},0)|^2$ as follows. We first draw a random variable $X \sim |\boldsymbol{\Psi}_{1}''|^2$ using the inversion method described above. The inverse cumulative function for this distribution is
\begin{equation}
F^{-1}(X) = s_{0}\sqrt{-2\ln(1-X)}.
\end{equation}
We then draw a random variable uniformly distributed on $(0,2\pi)$, $Y \sim \mathcal{U}(0,2\pi)$. We finally define
\begin{equation}
\left\{
\begin{aligned}
Z_{A} &= X\cos{Y}, \\
Z_{B} &= X\sin{Y},
\end{aligned}
\right.
\end{equation}
which yields $(Z_{A},Z_{B})$ distributed according to the modulus-squared of the wave function, as desired.

\subsection{Spin measurement}\label{spinmesurment}

So far we have focused on the spatial evolution of the particles. To analyze the Bell experiment, this spatial information must be related to the measurement outcomes recorded by the Stern-Gerlach detectors. In a Stern-Gerlach apparatus, spin information is converted into a spatial separation of the wave packet. The two spin components correspond to wave packets that move in opposite directions along the $z$ axis. At sufficiently long times these packets become well separated, so that the measurement outcome can be unambiguously inferred from the particle position. Within this framework, the measurement defines a mapping from particle position to spin outcome. We therefore associate the outcomes with the late-time particle positions according to

\begin{equation}
s_A=\mathrm{sgn}(z_A), \qquad s_B=\mathrm{sgn}(z_B).
\end{equation}

This identification completes the dynamical mapping from initial conditions to measurement outcomes.

\subsection{Trajectory patterns as a function of the relative angle $\gamma$}
\begin{figure}[hbtp]
\centering

\includegraphics[width=\columnwidth]{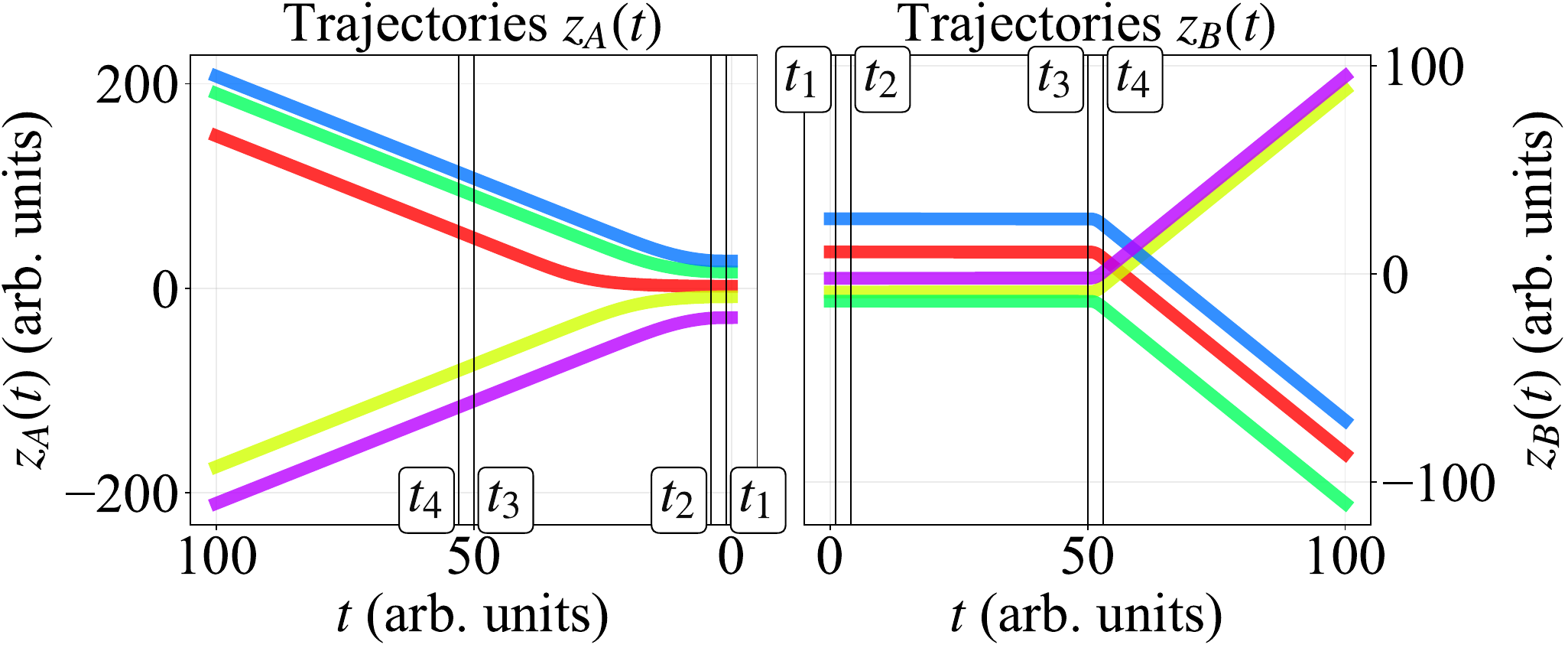}\par
{\small\centering $(a) \hspace{0.2cm} \gamma=0$\par}

\vspace{0.6em}

\includegraphics[width=\columnwidth]{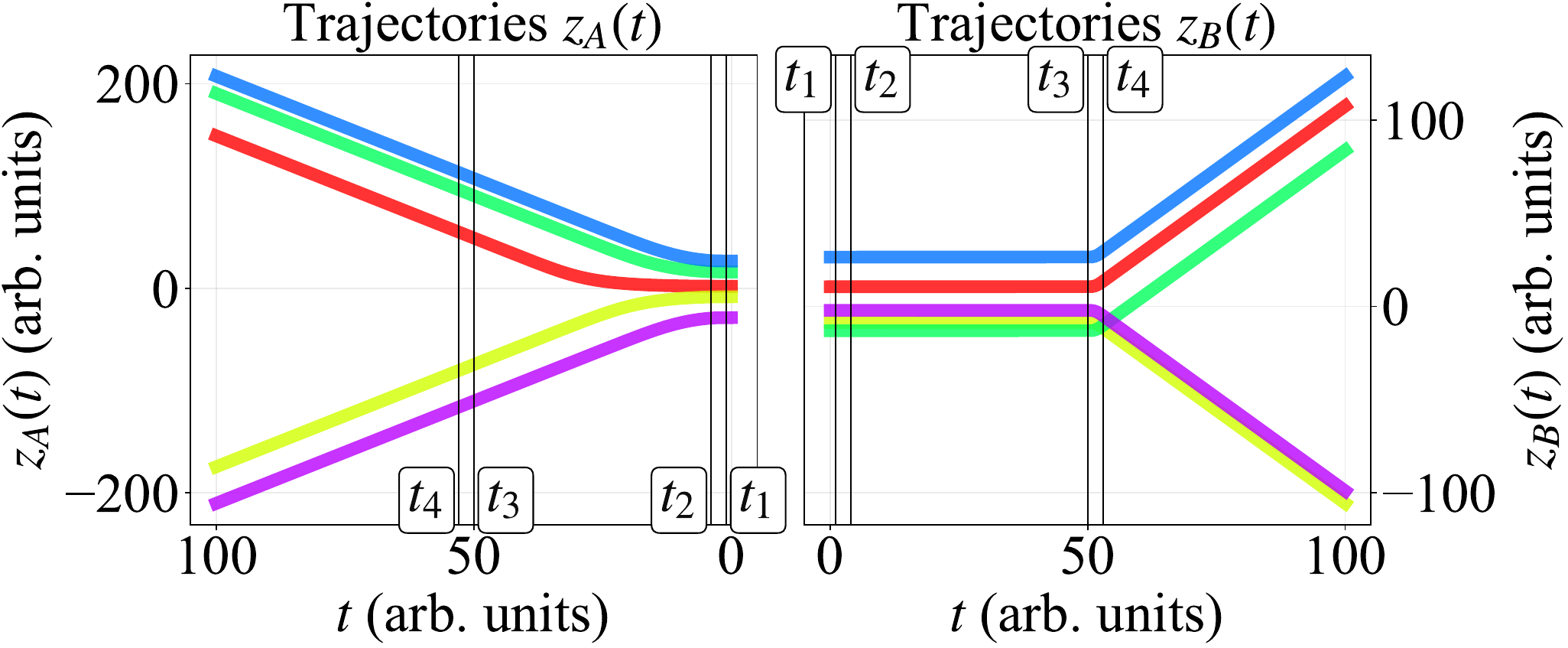}\par
{\small\centering $(b) \hspace{0.2cm} \gamma=\pi$\par}

\vspace{0.6em}

\includegraphics[width=\columnwidth]{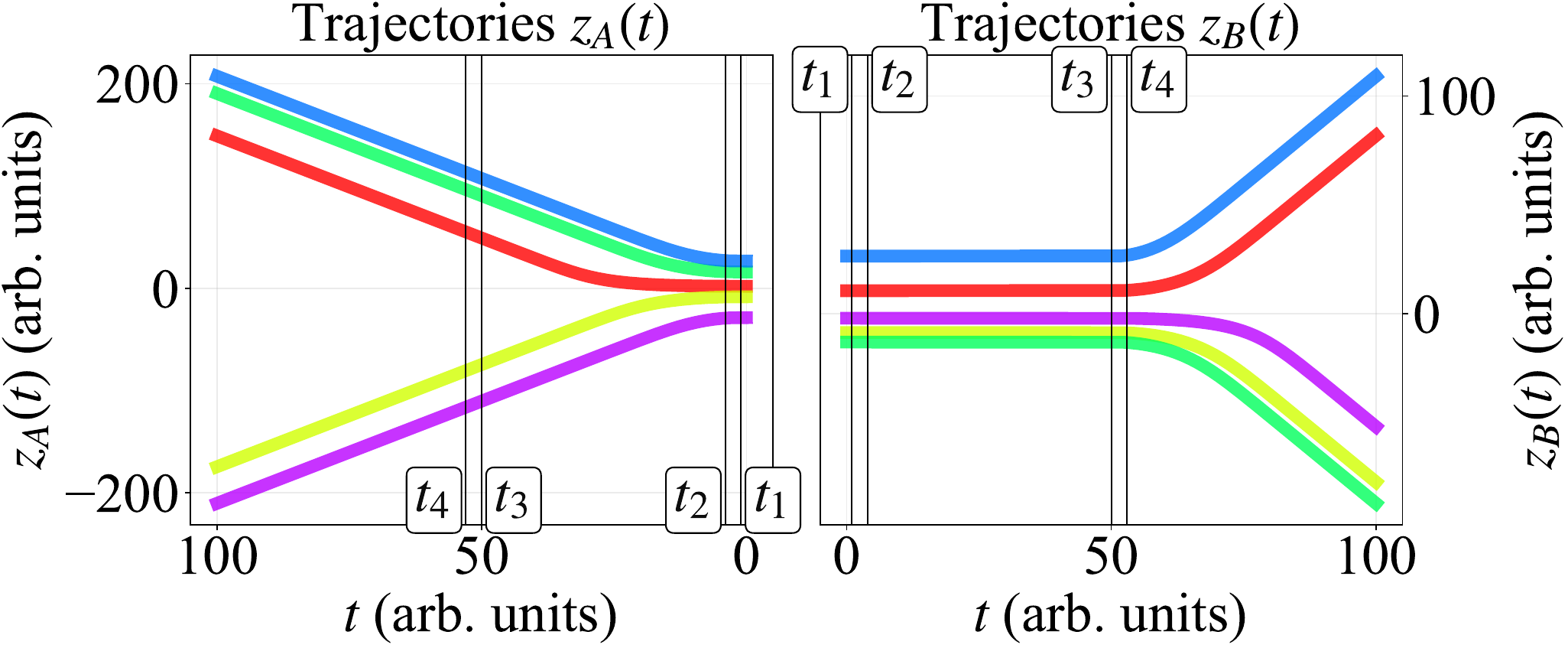}\par
{\small\centering $(c) \hspace{0.2cm}\gamma=\pi/2$\par}

\caption{\label{fig:trajectories_gamma}
Bohmian trajectories in the EPR-Bell pilot-wave model for three values of the relative coil angle $\gamma=\beta-\alpha$. Each panel displays the time evolution of $z_A(t)$ (left) and $z_B(t)$ (right) for Alice and Bob, respectively; trajectories corresponding to the same entangled pair are shown in the same color. The trajectories represent the Bohmian configuration-space dynamics of the particles, as represented through the measured spin coordinates. The resulting structure depends on the measurement configuration parameter $\gamma$, which encodes the relative coil settings. For $\gamma=0$ the dynamics yields perfect anticorrelation, for $\gamma=\pi$ perfect correlation, and for $\gamma=\pi/2$ an effectively factorized evolution of the trajectories. All panels are generated from the same ensemble of initial conditions, allowing a direct comparison of how the configuration-space dynamics reorganizes the trajectory structure as a function of $\gamma$.}
\end{figure}

We now examine the Bohmian trajectories associated with pairs of entangled particles for different values of $\gamma=\beta-\alpha$, shown in Fig.~\ref{fig:trajectories_gamma}. In each plot, the trajectories of particle $A$ $(z_{A}(t))$ are shown on the left for different initial conditions, while the trajectories of particle $B$ $(z_{B}(t))$ are shown on the right. For each realization, two particles are entangled, and the trajectories of an entangled pair are drawn with the same color.\\

\paragraph{Perfect anticorrelation: $\gamma=0$}\label{subsubsec:alphazéro}
We first consider the reference configuration $\gamma=0$, for which the two-particle state takes the singlet form and leads to perfect spin anticorrelation (see Fig.~\ref{fig:trajectories_gamma}(a)). The corresponding wave function reads : 
\begin{equation}
\boldsymbol{\Psi}_{4,\gamma=0}=\frac{\boldsymbol{\Psi}_{A,+} \boldsymbol{\Psi}_{B,-}-\boldsymbol{\Psi}_{A,-}\boldsymbol{\Psi}_{B,+}}{\sqrt{2}}.
\end{equation}
This wave function implies perfect spin anticorrelation. On the basis of this state, standard quantum mechanics predicts that if particle $A$ is measured to have spin $\pm$ then particle $B$ is necessarily measured to have spin $\mp$. The corresponding trajectory structure is consistent with this behavior.

Examining the trajectories more closely, we observe that some of them cross. For instance, if the initial positions satisfy $z_{0}>0$, particle $A$ is deflected upward and is interpreted as having spin $+1/2$. Particle $B$, which is subjected to the Stern-Gerlach field later, must then be deflected downward in order to be interpreted as having spin $-1/2$. To achieve this, it must cross the wave packet and reach the lower region of the screen, which implies an apparent crossing with other trajectories.

Such a crossing might appear problematic, since it seems to suggest that the same position could be associated with two different velocities. This apparent difficulty is resolved by noting that the dynamics takes place in configuration space: the state of the system is given by $(z_{A}(t),z_{B}(t),\vec v_{A}(t),\vec v_{B}(t))$, not by a single individual coordinate, with the velocities determined by the guidance equation. Two trajectories can therefore project onto the same point in physical space without violating the uniqueness of solutions in configuration space.\\

\paragraph{Perfect correlation: $\gamma=\pi$}\label{subsubsec:alphapi}

We now turn to the regime $\gamma=\pi$ (see Fig.~\ref{fig:trajectories_gamma}(b)), which can be viewed as the complementary situation to the singlet-like case discussed above. In this configuration the relative rotation induced by the coils effectively flips the correlation pattern: instead of perfect anti-correlation, the outcomes become perfectly \emph{correlated}. In other words, when particle $A$ is found in the upper (resp.\ lower) branch of its Stern-Gerlach analyzer, particle $B$ is subsequently guided toward the upper (resp.\ lower) branch as well. This behavior is clearly visible in the corresponding trajectory plot, where pairs of entangled trajectories now end up on the same side of their respective screens.

Here, the striking point is that the Bohmian dynamics on Bob’s side is not determined by local features of Bob’s apparatus alone: it depends on the experimental parameter $\gamma=\beta-\alpha$, and therefore on Alice's setting $\alpha$, even though Bob performs his measurement only later and can be arbitrarily far away. Changing the coil setting on Alice's branch modifies the global two-particle wave function in configuration space, and hence modifies the guidance field that determines Bob's velocity $\vec v_B=\vec j_B/\rho$. The fact that the correlation structure switches from anti-correlated ($\gamma=0$) to correlated ($\gamma=\pi$) by acting locally on Alice's side provides a direct illustration of the nonlocal character of the dynamics (and, more generally, of quantum mechanics): the effective guidance experienced by particle $B$ is instantaneously sensitive to distant experimental conditions. 

Comparing the cases $\gamma=0$ and $\gamma=\pi$ provides a direct illustration of the nonlocal character of the Bohmian dynamics. This behavior has been discussed qualitatively in the literature, notably by David Z. Albert \cite{Albert} and Jean Bricmont \cite{Bricmont}, and is here realized explicitly within the present dynamical framework.

\paragraph{Factorized dynamics: $\gamma=\tfrac{\pi}{2}$}

When the relative angle is $\gamma=\tfrac{\pi}{2}$, the underlying dynamical mapping does not induce correlations between the particle outcomes, as shown in Fig.~\ref{fig:trajectories_gamma}(c). Moreover, there are no crossings between trajectories, in this case, the two-particle dynamics reduces to two independent one-particle dynamics. Indeed, we have
\begin{equation}
\boldsymbol{\Psi}_{4,\frac{\pi}{2}}=\frac{1}{\sqrt{2}}\bigl(\boldsymbol{\Psi}_{A,+}( \boldsymbol{\Psi}_{B,-}+\boldsymbol{\Psi}_{B,+})+ \boldsymbol{\Psi}_{A,-}(\boldsymbol{\Psi}_{B,+}-\boldsymbol{\Psi}_{B,-})\bigr),
\end{equation}
so that
\begin{equation}
\begin{cases}
\displaystyle
\begin{split}
\vec{j}_{A/B}
&=\frac{\hbar}{2m}
\bigl(|\Psi_{B/A,+}|^2+|\Psi_{B/A,-}|^2\bigr)\\
&\qquad
\times\Im\!\Bigl(
\boldsymbol{\Psi}_{A/B,+}^\dagger
\vec{\nabla}_{A/B}\boldsymbol{\Psi}_{A/B,+} + \boldsymbol{\Psi}_{A/B,-}^\dagger
\vec{\nabla}_{A/B}\boldsymbol{\Psi}_{A/B,-}
\Bigr)
\end{split}
\\[1em]
\displaystyle
\rho
=\frac{1}{2}
\bigl(|\Psi_{A,+}|^2+|\Psi_{A,-}|^2\bigr)
\bigl(|\Psi_{B,+}|^2+|\Psi_{B,-}|^2\bigr).
\end{cases}
\end{equation}

and thus the velocities are :
\begin{equation}
\vec{v}_{A/B}=\frac{\hbar}{m}\Im\!\left(\frac{\boldsymbol{\Psi}_{A/B,+}^\dagger\vec{\nabla}\boldsymbol{\Psi}_{A/B,+}+ \boldsymbol{\Psi}_{A/B,-}^\dagger\vec{\nabla}\boldsymbol{\Psi}_{A/B,-}}{|\Psi_{A/B,+}|^2+|\Psi_{A/B,-}|^2}\right)
\end{equation}
These are precisely the velocity expressions for two independent one-particle dynamics. This behavior is consistent with the expected quantum correlations for this configuration.
 
\section{Bell inequalities and nonlocality} 

The previous section revealed strong correlations between the trajectories of the two particles, even though Alice and Bob can be arbitrarily far apart. While such correlations suggest nonlocality, establishing it requires showing that they cannot be reproduced by locally preexisting “elements of reality,” as introduced in the EPR argument \cite{epr}. Bell later formalized this requirement through statistical constraints -Bell inequalities- that any local hidden-variable theory must satisfy.

Since our model satisfies equivariance and is based on a standard Bell state, it reproduces the usual quantum mechanical statistics in quantum equilibrium. In our construction, these correlations arise from a deterministic mapping of initial conditions (hidden variables) to outcomes. It is therefore natural to test whether the induced statistics satisfy or violate Bell inequalities. We perform this test numerically by repeating the experiment many times, sampling initial conditions according to $|\Psi|^2$, and counting the resulting spin outcomes for different measurement settings.


\subsection{Bell-CHSH inequality}

Bell inequalities distinguish correlations compatible with local hidden-variable theories from those predicted by quantum mechanics. In a local model the outcomes of measurements performed by Alice and Bob, denoted $A(\alpha,\lambda)$ and $B(\beta,\lambda)$ with values $\pm1$, depend only on local settings ($\alpha$ or $\beta$) and on shared hidden variables $\lambda$. The average correlation is then
\begin{equation}
E(\alpha,\beta)=\int d\lambda\,\rho(\lambda)\,A(\alpha,\lambda)B(\beta,\lambda).
\end{equation}

In the present framework, this average arises from the distribution of initial conditions and the induced partition of hidden-variable configuration space. From these assumptions one obtains the Bell-CHSH combination
\begin{equation}
M=E(\alpha,\beta)-E(\alpha,\beta')+E(\alpha',\beta)+E(\alpha',\beta'),
\end{equation}
which satisfies the local bound
\begin{equation}
|M|\le2.
\end{equation}
Quantum mechanics predicts stronger correlations for entangled states, up to the Tsirelson bound $|M|\le2\sqrt2$, thereby excluding any local description.

\subsection{Bell-CHSH inequality in the dBB framework}

We now apply this reasoning to the outcome statistics induced by our dynamical construction. In our model the spin outcomes are inferred from the signs of the particle coordinates (see Sec.~\ref{spinmesurment}). We therefore define the binary observables
\begin{equation}
a_\alpha=\mathrm{sgn}(z_A)_\alpha, \qquad
b_\beta=\mathrm{sgn}(z_B)_\beta,
\end{equation}
where $\alpha$ and $\beta$ denote the rotation angles induced by the coils.

The Bell variable is then
\begin{equation}
S(\alpha,\alpha',\beta,\beta')=
a_\alpha b_\beta-a_{\alpha'} b_\beta
+a_\alpha b_{\beta'}+a_{\alpha'} b_{\beta'},
\end{equation}
and its average
\begin{equation}
M=\langle S\rangle
\end{equation}
must satisfy $|M|\le2$ for any local theory.

To compare with the quantum prediction we consider
\begin{equation}
M(\theta)=\langle S(0,\theta,\tfrac{\theta}{2},\tfrac{3\theta}{2})\rangle ,
\end{equation}
for which
\begin{equation}
M(\theta)=3\cos(\tfrac{\theta}{2})-\cos(\tfrac{3\theta}{2}).
\end{equation}
 
We estimate $M(\theta)$ numerically by sampling initial conditions according to $|\Psi|^2$, evolving the corresponding trajectories, and extracting the associated outcomes. Increasing the number of simulated particle pairs reduces the statistical fluctuations of the estimate $M^{\text{num}}(\theta)$.

\begin{figure}
    \centering
    \includegraphics[width=1\columnwidth]{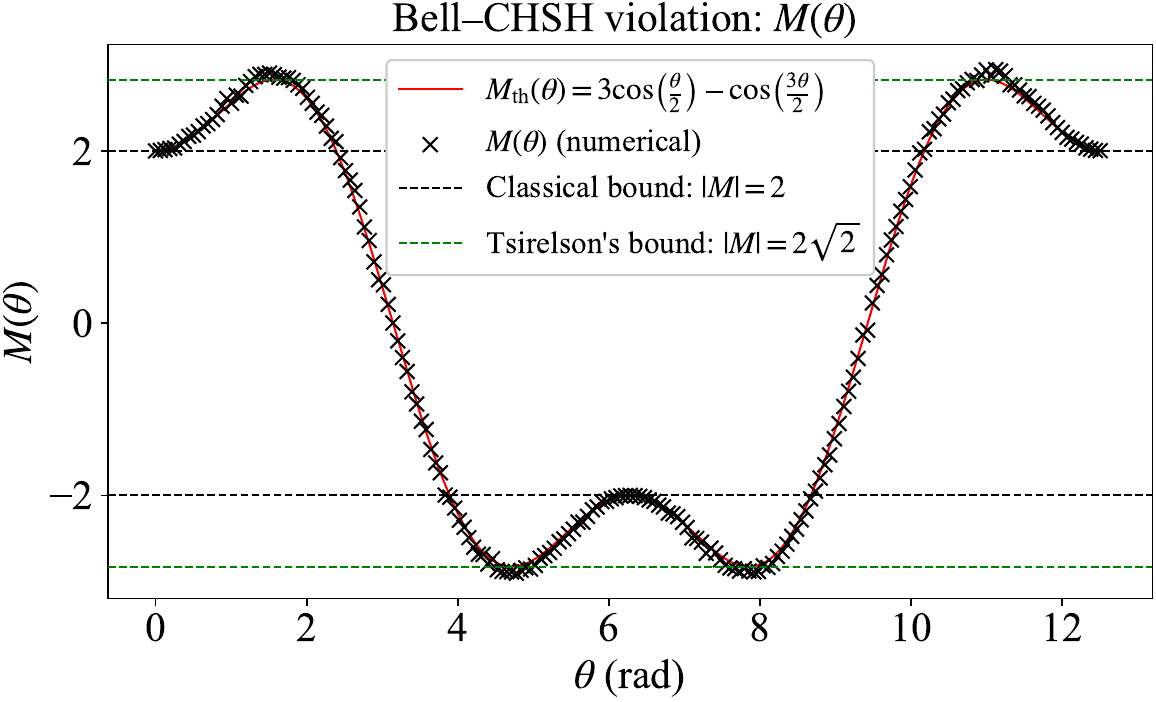}
    \caption{\label{fig:bell}
Bell-CHSH parameter for the EPR-Bell pilot-wave model. For each value of $\theta$, the correlation combination $M(\theta)=\langle S(0,\theta,\tfrac{\theta}{2},\tfrac{3\theta}{2})\rangle$ is computed from Bohmian trajectories by extracting binary outcomes $a_\alpha=\mathrm{sgn}(z_A)$ and $b_\beta=\mathrm{sgn}(z_B)$ at late times. Crosses: numerical estimates obtained from $200$ values of $\theta$ with $2000$ entangled pairs per value (quantum-equilibrium sampling). Solid curve: quantum prediction $M(\theta)=3\cos(\tfrac{\theta}{2})-\cos(\tfrac{3\theta}{2})$. This quantity provides a consistency check of the configuration-space construction, confirming that the induced hidden-variable dynamics reproduces the expected Bell-CHSH correlations. The region $|M(\theta)|>2$ signals violation of the local bound, with a maximum consistent with the Tsirelson limit $2\sqrt{2}$.}
\end{figure}

Figure~\ref{fig:bell} shows that the numerical values closely follow the theoretical curve. The deviations from the curve are consistent with statistical fluctuations due to the finite sample size (2000 pairs per value of $\theta$). The region where $|M(\theta)|>2$ clearly demonstrates the violation of the Bell-CHSH bound, confirming that the Bohmian dynamics reproduces the expected quantum correlations.

\subsection{Illustration of the no-signaling theorem}

We now examine how no-signaling emerges within the present construction, despite the explicit nonlocal dependence of the dynamics. The no-signaling property ensures compatibility with special relativity, which prohibits faster-than-light communication. In the present framework, although the outcome mapping depends nonlocally on the global measurement configuration, the local statistics remain invariant under changes of distant settings.

The no-signaling property was established rigorously in the late 1970s \cite{Bellnote,Eberhard,Ghirardi}, although its basic intuition was already present in earlier EPR-Bell discussions. The central point is that, while entangled particles exhibit strong nonlocal correlations, these correlations cannot be exploited in a controlled way to influence the marginal statistics observed by a distant observer.

In the context of our experiment, consider that Alice modifies the orientation $\alpha$ of her coil in an attempt to influence Bob’s results. If superluminal communication were possible, Bob’s local statistics would depend on Alice’s choice. However, this does not occur. Although the \emph{joint} outcomes depend on the relative angle $\gamma=\beta-\alpha$, the \emph{marginal} statistics measured by each observer remain unchanged: Bob always observes equal proportions of spin-up and spin-down outcomes, independently of Alice’s setting (and vice versa). The dependence on the distant setting appears only at the level of correlations, revealed when the data are compared through a classical communication channel.

To illustrate this mechanism within our numerical model, we introduce a representation of the hidden-variable configuration space. In Bohmian mechanics, each realization of the experiment is fully determined by the initial particle positions. By representing these initial conditions as points on a unit disk and assigning to each point the corresponding measurement outcomes, we obtain a direct visualization of how the hidden-variable space is partitioned into outcome domains. While the geometry of these domains depends on the measurement settings, their projected measures associated with local outcomes remain unchanged, thereby providing a concrete illustration of the no-signaling property.

\subsubsection{Hidden-variable representation of outcomes}

In our EPR-Bell-type experiment, the initial conditions are generated as follows.
We first draw a point $(r,\theta)$ uniformly on the unit disk. This point is then transformed to obtain the initial conditions $(z_A(t=0),z_B(t=0))$ that determine the dynamics. The transformation, via the bijection $F$, is
\begin{equation}
\begin{gathered}
(z_A(t=0),z_B(t=0))
=F(r,\theta)
\\[4pt]
=s_{0}\sqrt{-2\ln(1-r)}\,(\cos\theta,\sin\theta).
\end{gathered}
\end{equation}
Sampling the initial conditions according to $|\boldsymbol{\Psi}_{1}(z_A,z_B,0)|^2$ is thus equivalent to uniformly sampling points on the unit disk. These points parameterize the hidden-variable configuration space and fully determine the outcome through the deterministic dynamics. They are equivalent to the particle positions, since $F$ is a bijection between the disk and $\mathbb{R}\otimes\mathbb{R}$, the space of particle positions.

Quantum equilibrium then corresponds to a uniform distribution on this disk: drawing a point uniformly on the disk and transforming it via $F$ is equivalent to sampling initial conditions according to $|\boldsymbol{\Psi}_{1}(z_A,z_B,0)|^2$. Since pilot-wave theory is deterministic, each set of initial conditions (i.e., each pair of hidden variables) is associated with a unique outcome of the experiment.

In the next section, we (randomly) place points representing initial conditions for our EPR-Bell experiment on the unit disk. We then solve the dynamics for each of these points, exactly as in the preceding sections. We color each point according to the outcome associated with the initial condition that it represents.

\begin{figure}
\centering
\begin{minipage}{0.49\columnwidth}
  \centering
  \includegraphics[width=\linewidth, clip, trim=0cm 2.7cm 0cm 1.5cm]{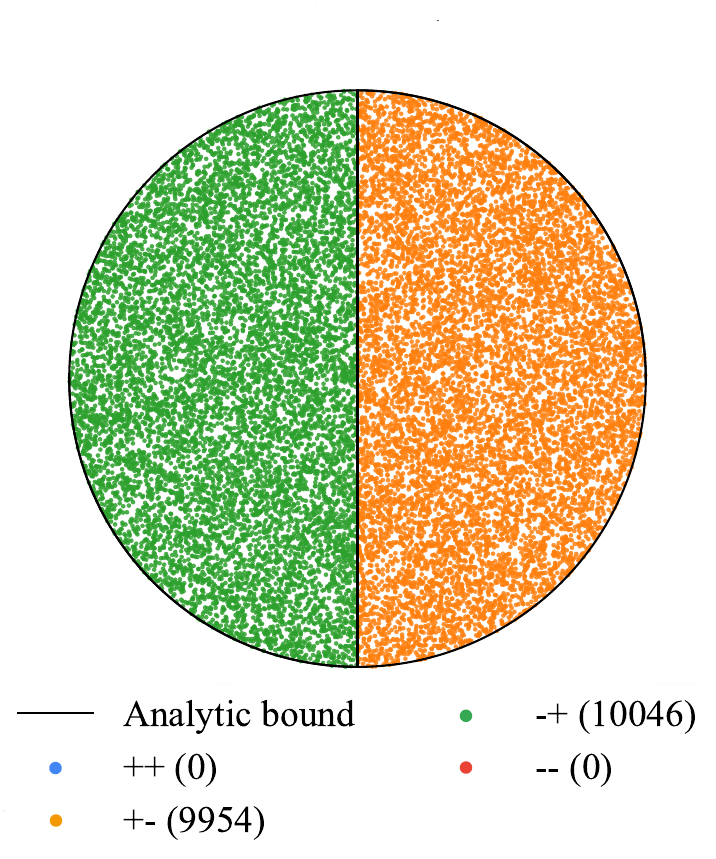}\\[-0.3em]
  {\small (a) $\gamma=0$}
\end{minipage}\hfill
\begin{minipage}{0.49\columnwidth}
  \centering
  \includegraphics[width=\linewidth, clip, trim=0cm 2.7cm 0cm 1.5cm]{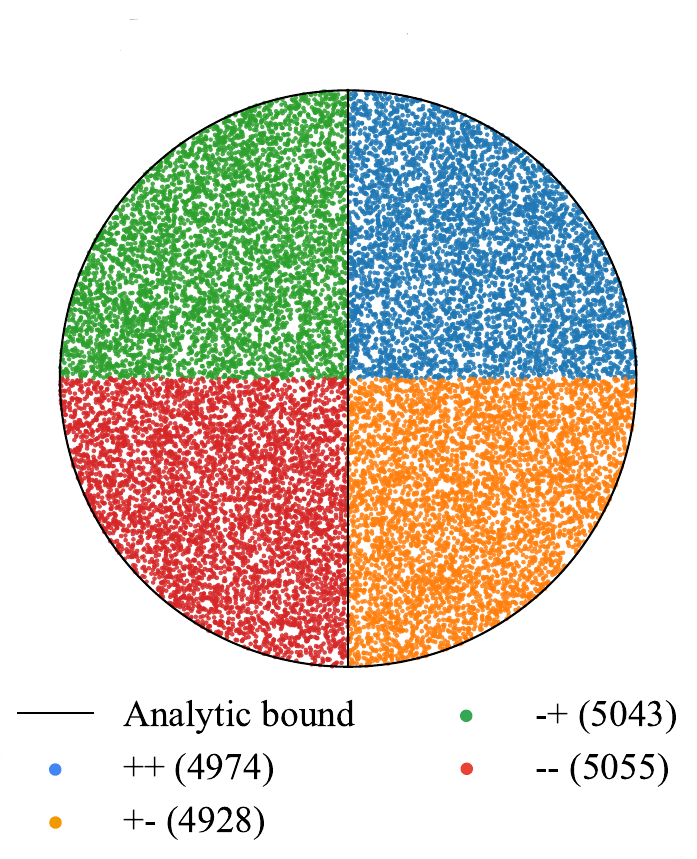}\\[-0.3em]
  {\small (b) $\gamma=\pi/2$}
\end{minipage}

\vspace{0.6em}

\begin{minipage}{0.49\columnwidth}
  \centering
  \includegraphics[width=\linewidth, clip, trim=0cm 2.7cm 0cm 1.5cm]{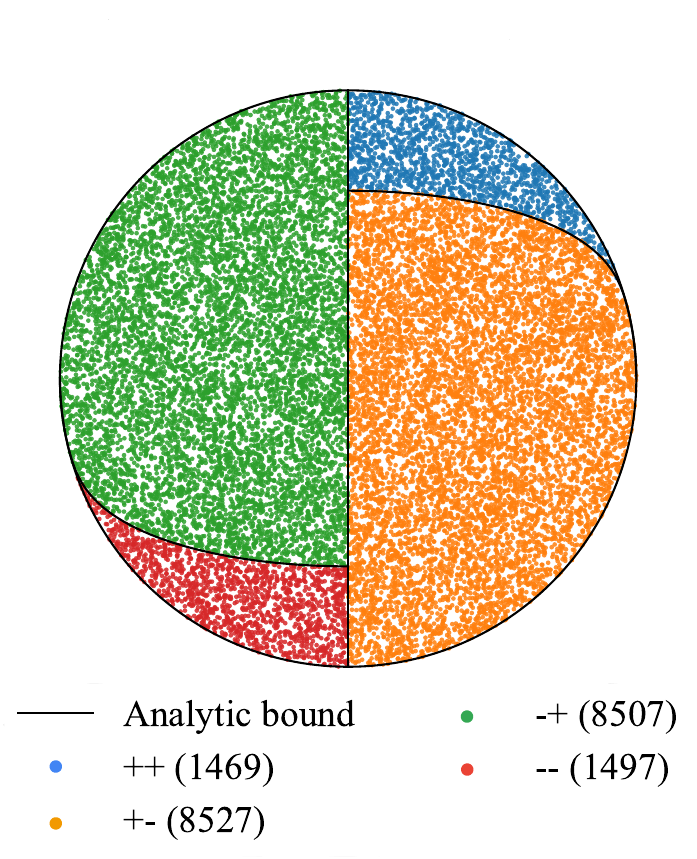}\\[-0.3em]
  {\small (c) $\gamma=\pi/4$}
\end{minipage}\hfill
\begin{minipage}{0.49\columnwidth}
  \centering
  \includegraphics[width=\linewidth, clip, trim=0cm 2.7cm 0cm 1.5cm]{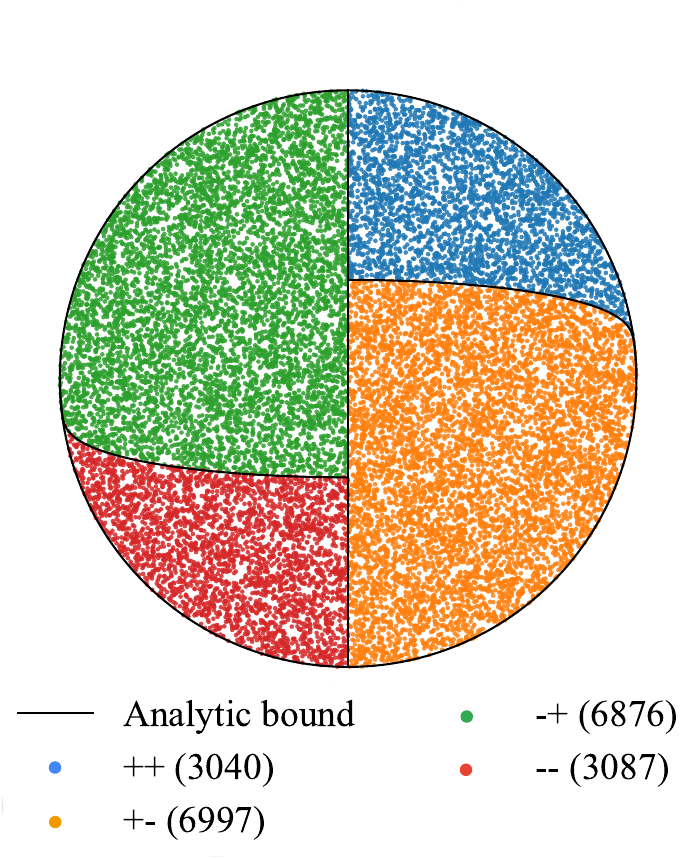}\\[-0.3em]
  {\small (d) $\gamma=3\pi/8$}
\end{minipage}
\caption{\label{fig:disks_outcomes}
Hidden-variable (unit-disk) representation of outcomes in the EPR-Bell pilot-wave model for four relative coil angles $\gamma = \beta - \alpha$. Each point corresponds to one initial condition, uniformly sampled on the unit disk (quantum equilibrium) and mapped to $(z_A(0), z_B(0))$ via the bijective transformation defining the configuration-space representation of the model. Colors encode the four joint outcomes $(s_A, s_B) \in {+,-}^2$, obtained from the late-time signs $s_A = \mathrm{sgn}(z_A)$ and $s_B = \mathrm{sgn}(z_B)$: blue for $(+,+)$, orange for $(+,-)$, green for $(-,+)$, and red for $(-,-)$. The induced partition of the hidden-variable configuration space depends on the measurement configuration parameter $\gamma$, which encodes the relative coil settings. The resulting domain structure evolves with $\gamma$: the partition boundaries reflect the dynamical dependence of the outcome mapping on the measurement settings, while preserving setting-independent marginal statistics. The solid line represents the analytically determined separatrix between outcome domains, and the numerically sampled regions coincide with it (within numerical resolution), confirming the predicted configuration-space structure. Although the geometry of the partition depends nonlocally on $\gamma$, the marginal distributions remain balanced for all settings, $P_A(+) = P_A(-) = \tfrac{1}{2}$ and $P_B(+) = P_B(-) = \tfrac{1}{2}$, ensuring no-signaling at the statistical level.}
\end{figure}

This defines a partition of the hidden-variable space into four outcome domains $(+_A,+_B)$, $(+_A,-_B)$, $(-_A,+_B)$, and $(-_A,-_B)$
by coloring the points associated with the corresponding initial conditions (see Fig.~\ref{fig:disks_outcomes}).\\
\paragraph{Mathematical analysis}

On our disk of initial conditions, we expect to see four distinct sectors corresponding to the different outcomes $(+_A,+_B)$, $(+_A,-_B)$, $(-_A,+_B)$ and $(-_A,-_B)$. Our aim here is to predict the shapes of the boundaries between these differently colored sectors.\\
First recall that, in our model, Bob performs his measurement only much later than Alice. Thus, at the moment particle $B$ enters its Stern-Gerlach device, particle $A$ is already far from the central axis and its spin is determined with certainty. This induces a vertical boundary on the disk separating the regions $z_{A,0}<0$ and $z_{A,0}>0$. The spin measurement outcome for particle $A$ is independent (because sufficiently earlier) of the spin measurement outcome for particle $B$. Consequently, only the initial position of particle $A$ fixes its spin. By symmetry and as explained above, if $z_{A,0}>0$ (resp.\ $z_{A,0}<0$), then particle $A$ will be spin up (resp.\ spin down).\\
We have thus identified the boundary separating $(-_A,\dots)$ from $(+_A,\dots)$. We now need the expressions for the two other boundaries: the one separating $(+_A,+_B)$ from $(+_A,-_B)$, and the one separating $(-_A,-_B)$ from $(-_A,+_B)$. The spin outcomes for particle $B$ are likewise entirely determined by the initial conditions. We first examine the marginal probability density for spin $B$ given that the spin of $A$ is positive.\\
This density is given by
\begin{equation}
\frac{\pi_{\left|+_A\right\rangle}(\boldsymbol{\Psi}_{4})}{\|\pi_{\left|+_A\right\rangle}(\boldsymbol{\Psi}_{4})\|},
\end{equation}
which is equal to :
\begin{equation}
\sin(\gamma/2)\boldsymbol{\Psi}_{+}(z_B,t)+\cos(\gamma/2)\boldsymbol{\Psi}_{-}(z_B,t).
\end{equation} 
This shows that the marginal probability density associated with particle $B$ reduces to an asymmetric Stern-Gerlach density of the form \eqref{spinsg1}:
\begin{equation}
c_+\boldsymbol{\Psi}_{+}(z_B,t)+c_-\boldsymbol{\Psi}_{-}(z_B,t),
\end{equation} 
with $c_+=\sin(\gamma/2)$ and $c_-=\cos(\gamma/2)$. Observing that we are dealing with an asymmetric one-particle Stern-Gerlach experiment, and knowing that the fraction $|c_+|^2$ of $B$ particles with the largest initial $z_{B,0}$ values will be spin up, we can deduce the boundary separating the $(+_A,+_B)$ and $(+_A,-_B)$ domains.
Indeed, the coordinate $z_{B,0}$ marking the boundary between the $(+_A,-_B)$ and $(+_A,+_B)$ regions is determined by
\begin{equation}
\int_{z_{B,0}}^{+\infty} \frac{1}{\sqrt{2\pi} s_0}e^{-\frac{z^2}{2s_0^2}}dz=|c_+|^2=\sin^{2}(\gamma/2).
\end{equation}
Solving the integral and inverting to find $z_{B,0}$, we obtain
\begin{equation}
z_{B,0}=\sqrt{2}\,s_0\,\operatorname{erf}^{-1}(\cos(\gamma)),
\end{equation}
where $\operatorname{erf}$ is the error function. A constant abscissa $z_{B,0}$ in real space corresponds to a parametric curve $U(\theta)$ on the disk. Indeed,
\begin{equation}
z_{B,0}=r\sin(\theta)=s_0\sqrt{-2\ln(1-U)}\sin(\theta),
\end{equation}
and thus
\begin{equation}
\begin{gathered}
U(\theta)=1-\exp\!\left[-\left(\frac{\operatorname{erf}^{-1}(\cos\gamma)}{\sin\theta}\right)^{2}\right],
\\[4pt]
\theta \in \left[0;\frac{\pi}{2}\right].
\end{gathered}
\end{equation}
Repeating the same reasoning to find the boundary between the $(-_A,+_B)$ and $(-_A,-_B)$ domains, we obtain the same formula for $\theta \in \left[\pi;\frac{3\pi}{2}\right]$.
We can thus verify analytically that our boundaries on the disk between the different regions coincide with the expected results.

We observe a clear influence of the relative angle induced by the coils on the shape of the domains on the disk. However, the only data available to Alice and Bob are the proportions of spin-up and spin-down outcomes on their respective branches. Remarkably (and as expected), Alice and Bob always measure equal fractions of spin-up and spin-down, whatever the angle induced by the coils. It is therefore impossible for Bob to detect a change in the angle induced by Alice's coil. This illustrates the no-signaling property: although the configuration-space partition of hidden variables depends nonlocally on the measurement settings, the resulting marginal statistics remain invariant.

\section{Conclusion}

We have analyzed an explicit EPR-Bell construction within de Broglie-Bohm theory, showing how the measurement dynamics induces a partition of hidden-variable configuration space into domains associated with distinct joint outcomes. Using a reduced-dimensional Stern-Gerlach model with local spin rotations, we analyzed how the measurement dynamics induces a partition of hidden-variable configuration space into domains associated with distinct joint outcomes. The choice of a one-dimensional model, in which the spatial dynamics is carried by Gaussian wave packets propagating along the $z$ axis, allows us to maintain a simple geometry while incorporating spin, entanglement, and nonlocality explicitly.

In the single-particle case, the analysis of the Stern-Gerlach experiment showed how the wave function splits into two branches corresponding to the ``$+$'' and ``$-$'' spin components. Pilot-wave theory then provides a deterministic description: the trajectory of a particle is fully determined by its initial position through the guidance equation, and the statistics of outcomes ($|c_\pm|^2$) are recovered when initial conditions are distributed according to $|\Psi|^2$. This emphasizes that the measured spin value is not a preexisting binary property, but emerges from the combined effect of the wave function structure and the initial condition.

Extending to the bipartite case, we constructed an antisymmetric Bell state and analyzed the dynamics of two entangled particles. Local spin rotations, implemented through interaction Hamiltonians in magnetic coils, play the role of the relative orientations of the Stern-Gerlach analyzers in Bell’s original setup. We identified characteristic regimes as a function of the relative angle $\gamma = \beta-\alpha$: perfect anticorrelations for $\gamma=0$, two effectively independent dynamics for $\gamma=\pi/2$, and intermediate behavior for other values.

At the statistical level, we defined binary observables $a_\alpha$ and $b_\beta$ from the signs of the final coordinates $z_A$ and $z_B$, and constructed the Bell-CHSH combination
\begin{equation}
M(\theta)=  \langle S(0,\theta,\tfrac{\theta}{2},\tfrac{3\theta}{2}) \rangle.
\end{equation}

Numerical simulations, based on quantum-equilibrium sampling ($\rho=|\boldsymbol{\Psi}|^2$) and integration of the guidance equations, show that $M^{\mathrm{num}}(\theta)$ closely follows the analytical prediction
\begin{equation}
M(\theta)=3\cos\!\left(\frac{\theta}{2}\right)-\cos\!\left(\frac{3\theta}{2}\right),
\end{equation}
and violates the local bound $|M|\leq 2$ up to the Tsirelson limit $|M|\leq 2\sqrt{2}$. These results provide an explicit consistency check of how deterministic hidden-variable dynamics with intrinsic nonlocality reproduces the statistical signature of EPR-Bell correlations.

\indent Finally, we analyzed the no-signaling property within the same framework. By representing initial conditions on the unit disk, in bijection with the configuration space $(z_A,z_B)$, and coloring each point according to the outcome $(s_A,s_B)$, we obtained a partition of hidden-variable configuration space whose geometry depends on the relative angle $\gamma$. This dependence reflects the nonlocal structure of the Bohmian dynamics: changes in Bob’s measurement setting modify the global guidance field in configuration space. However, the marginal distributions of local outcomes (the proportions of $s_A=\pm1$ and $s_B=\pm1$) remain independent of distant settings, in agreement with the no-signaling constraint.

Several natural extensions of this work can be envisaged. From a numerical standpoint, it would be interesting to refine the treatment of the magnetic field regions by integrating the Pauli equation directly in the interaction zone, rather than using an effective interpolation between free regimes. On the physical side, extending the model to three dimensions would bring it closer to actual experimental setups and allow exploration of arbitrary analyzer orientations.

Finally, the formalism presented here could be used to investigate quantum non-equilibrium situations in which the initial distribution departs from $\rho=|\boldsymbol{\Psi}|^2$. Such simulations would illustrate how, outside equilibrium, Bohmian nonlocality can in principle lead to signaling effects, thereby providing a sharp contrast between standard quantum equilibrium and hypothetical non-equilibrium regimes \cite{Valentinia,Valentinib}.\\



{\bfseries ACKNOWLEDGEMENTS}\\

\noindent We would like to thank Travis Norsen, Jean Bricmont and Siddhant Das for their valuable feedback and suggestions. We also thank Klaus M\o lmer for sharing related but independent work carried out together with Viktor K. Haldborg that is not yet published.\\

{\bfseries DATA AVAILIBILITY STATEMENT}\\

\noindent  The data generated in this study are available from the authors upon reasonable request.


\begin{thebibliography}{99}
\bibitem{note1}
DBB theory is often referred to as Bohmian mechanics in the literature \cite{DurrTeufel,DurrLazarovici}.
\bibitem{note2}
Strictly speaking, what is violated is the joint property of local causality and the absence of `superdeterminism'. The latter is generally considered unlikely. Similarly, we note that in the literature, the non-consensual term impossiblity  of `local-realism' is sometimes used to refer to Bell's theorem, implying that quantum mechanics could ultimately be both local and non-realistic. Einstein was clearly a local-realist but neither realism nor determinism is supposed to derive Bell's theorem. In the following, we never use this terminology, which has been rightly criticized elsewhere (see \cite{Norsen,Norsen2,Maudlin,Laudisa,Drezetchapter}).
\bibitem{warning}
The influence of the spin current on trajectories is discussed in Refs. \cite{Ohanian,Mita,Hodge,Philippidis}. The empirical equivalence with regard to quantum statistics is subject to a recent controversy related to the notion of arrival time \cite{Das,GTZ,Drezettime}, which we will not address here.
\bibitem{noteRauch}
It should be noted that this type of spin-flipper analysis played a role in experiments analyzing quantum measurement theory  and Bohr's complementarity  in a Bohmian framework \cite{Vigier,Badurek,Vigier1}.
\end{thebibliography}
\end{document}